\documentstyle{elsart}

\def\Frac#1#2{\frac{\displaystyle{#1}}{\displaystyle{#2}}} 
\input{epsf}

\begin{document}
\begin{frontmatter}

\address{~~~~~~~~~~~~~~~~~~~~~~~~~~~~~~~~~~~~~~~~~~~~~~~~~~~~~
~~~~~~~~~~~~~~SISSA 39/2000/EP}

\title{Cosmology of the Randall-Sundrum model after dilaton stabilization}

\author{Julien Lesgourgues, Sergio Pastor, Marco Peloso, Lorenzo Sorbo}

\address{\it SISSA--ISAS, Via Beirut $2$-$4$, $34013$ Trieste, Italy\\ and \\
INFN, Sezione di Trieste, Via Valerio $2$, $34127$ Trieste, Italy}

\begin{abstract}
We provide the first complete analysis of cosmological evolution in the Randall-Sundrum model with stabilized dilaton. We give the exact expansion law for matter densities on the two branes with arbitrary equations of state. The effective four-dimensional theory leads to standard cosmology at low energy. The limit of validity of the low energy theory and possible deviations from the ordinary expansion law in the high energy regime are finally discussed.

\vskip 0.5cm

\end{abstract}

\end{frontmatter}

\section{Introduction}

The past two years have witnessed the rise of several proposals for solving the
hierarchy problem with the aid of large extra dimensions. In the original
suggestion of Arkani-Hamed {\it {et al.}} \cite{ADD}, the discrepancy between
the effective four-dimensional Planck mass and the electroweak scale originates
from the largeness of the extra dimensions in which only gravity can propagate.
However, it is not obvious that such a large volume can appear naturally. An
alternative scheme was proposed by Randall and  Sundrum  \cite{RS}, and is
based on so-called ``warped compactification''. In this model, two branes are
embedded into an anti-De-Sitter five-dimensional space-time (the bulk), and all
the mass parameters of the five-dimensional action are approximately of the
same order of magnitude. However, for moderately large values of the
compactification radius, a strong hierarchy appears between the effective
gravitational scale and the other mass scales in one of the branes. 

Large extra dimensions could in principle have a large impact on the
cosmological evolution of the Universe. This was first realized by Bin\'etruy
{\it {et al.}} \cite{BDL}, who showed that in a two-brane model, with an empty
bulk, the solution of the five-dimensional Einstein equations leads to a
phenomenologically unacceptable expansion law, with the Hubble parameter
proportional to the energy density $\rho$ in the branes (rather than to
$\sqrt{\rho}$ as in the Friedmann equation). This result changes in presence of
cosmological constants \cite{CGS,BDEL,SMS,FTW}. In particular, the Friedmann
equation is recovered at low energy, provided that the cosmological constants
in the bulk and on the branes are related as in \cite{RS}. However, in all
these works, a specific constraint between the energy densities of matter on
the branes follows from the Einstein equations, and leads to  negative energy
density in our brane, at odds with phenomenology.

Later on, Cs\'aki {\it {et al.}} \cite{CGRT} noticed that in presence of a
mechanism for the stabilization of the dilaton (or, more simply, of the
physical distance between the two branes), no constraint appears between the
energy densities on the two branes, and the standard expansion is automatically
recovered. However, in ref. \cite{CGRT} this was proved only for small energy
densities constant in time. Here, we want to include realistic matter on the
branes (with arbitrary equations of state) and to give a solution valid at high
energy, provided that the dilaton is already stabilized. We are especially
interested in understanding whether deviations from the standard expansion can
occur at some energy.

To achieve this goal, we will first give the exact solution of the
five-dimensional Einstein equations (section I). This derivation is obtained in
analogy with the one of \cite{KKOP} performed in the case of a single brane
(see also \cite{KIM}, where some of our results have also been achieved). Then,
in order to correctly interpret these results in terms of observable quantities
in our brane, we will calculate in section II the gravitational action of the
effective four-dimensional theory. In section III, we will show that at low
energy the effective theory can be easily defined, and no deviation from
standard four-dimensional cosmology can be observed. In particular, as in the
case considered in \cite{CGRT}, any kind of matter on one brane behaves exactly
as dark matter on the other.  Finally, in section IV, we will discuss the
evolution of the system at high energy, and show that no deviation is likely to
be produced when the energy density in our brane is smaller than (TeV)$^4$.

\section{Exact solutions in presence of matter}

We consider the Randall-Sundrum (RS) model with two branes and the metric:
\begin{eqnarray}
ds^2 &=& g_{AB} dx^A dx^B, \qquad
\mbox{\scriptsize (A,B) $\in$ \{0,1,2,3,5\} }
\nonumber \\
&=& n(y,t)^2 dt^2 - a(y,t)^2 d \vec{x}^2 - b_0^2 dy^2
\; .
\label{metric}
\end{eqnarray}
The variable $y$ parametrizes the extra dimension compactified on the interval $\left[ -\,1/2\,,\,1/2 \right]\,$. The two branes are located at $y=0$ and at $y=1/2\,$, and the $Z_2$ symmetry $y \leftrightarrow - \, y$ is imposed. 
We have assumed that the dilaton $b(y,t)$ has already been stabilized at the
value $b_0\,$. Since we are interested in the cosmological evolution after
this has occurred, we do not enter here in the details of the stabilization
mechanism (see however \cite{GW}).
The nontrivial components of the Einstein tensor associated to the metric (\ref{metric}) are:
\begin{eqnarray}
G_{00} &=& 3 \left( \frac{\dot{a}}{a} \right)^2
- 3 \frac{n^2}{b_0^2} \left[ \frac{a''}{a} 
+ \left( \frac{a'}{a} \right)^2 \right] \;,
\label{Einstein00} \\
G_{ii} &=& \frac{a^2}{n^2} \left[ -  \left( \frac{\dot{a}}{a} \right)^2
+ 2 \frac{\dot{a}}{a} \frac{\dot{n}}{n} - 2 \frac{\ddot{a}}{a} \right]
+ \frac{a^2}{b_0^2}
\left[ \left( \frac{a'}{a} \right)^2 + 2 \frac{a'}{a} \frac{n'}{n}
+ 2 \frac{a''}{a} + \frac{n''}{n} \right] \;,
\label{Einsteinii} \\
G_{05} &=& 3 \left[ \frac{n'}{n} \frac{\dot{a}}{a} - \frac{\dot{a}'}{a} \right] \;,
\label{Einstein05}
\end{eqnarray}
where dot denotes differentiation whith respect to $t$ and prime with respect to $y\,$.
We do not consider the component $G_{55}$, because the corresponding Einstein
equation accounts for the stabilization of the dilaton.\footnote{One
customarily assumes that a potential $U \left( b \right)$ is generated in
the $5$ dimensional theory by some mechanism \cite{GW}. If the dilaton is
very heavy, that is if near the minimum $\bar{b}$ we have $U \left(
b \right) \simeq M_b^5 \left( b - \bar{b} \right)^2 / \bar{b}^2$ with a
very high mass scale $M_b\,$, the $55$ Einstein equation gives $b_0 \simeq
\bar{b}$ plus small corrections proportional to the energy densities on
the two branes. If instead the dilaton is required to be fixed without any
stabilization mechanism (that is $U \left( b \right) = 0\,$), the $55$ 
equation forces a precise relation  between the energy densities on the
two branes, which requires the matter energy density on our brane to be negative
\cite{CGS,BDEL,SMS}.} In the RS scenario one introduces
cosmological constants both in the bulk ($\Lambda$) and on the two branes
($V_0$, $V_{1/2}$). These energies satisfy the relation:
\begin{equation}
V_0 = - \, V_{1/2} = -\,\frac{\Lambda}{m_0} = \frac{6 m_0}{\kappa^2} \;\;,
\label{finetun}
\end{equation}
where $\kappa^2$ is the proportionality factor in the Einstein equations
$
G_{AB} = \kappa^2 \, T_{AB},
$
and $m_0$ a mass parameter.\footnote{The relation (\ref{finetun}) is
equivalent to fine--tune to zero the cosmological constant in ordinary
Friedmann--Robertson--Walker (FRW) cosmology.}
With these choices, the system admits the solution:
\begin{equation}
n \left( y \right) = a \left( y \right) =e^{- m_0 b_0 |y|} \;\;.
\label{statsol}
\end{equation}
This solution gives flat space--time on the two branes. In particular,
Minkowski space--time (with a canonical four-dimensional action) is
recovered also in the $1/2-$brane after the redefinition of the
fields:
\begin{equation}
\phi \rightarrow \phi / \Omega_0  \;\;\;,\;\;\; \Omega_0 \equiv e^{-m_0 b_0 /2} \;\;.
\end{equation}
This redefinition modifies all the mass scales $m$ of the lagrangian of the $1/2-$brane, according to 
\begin{equation}
m \rightarrow \Omega_0 \, m \;\;.
\label{mred}
\end{equation}
This may provide a solution of the hierarchy problem if we assume that we live on the $1/2-$brane. To see this, one first notices from the five dimensional gravitational action that the four dimensional effective Planck mass is here given by
\begin{equation}
M_p^2 = \frac{1 - \Omega_0^2}{\kappa^2 \, m_0} \;\;.
\label{mpl}
\end{equation}
It thus appears natural to take all the mass scales of the system to
be of the order of the observed Planck mass. The measured smallness of
the electroweak scale follows then from the redefinition (\ref{mred})
of the masses on our brane. Since $\Omega_0$ depends exponentially on
the product $m_0 \, b_0\;$, it is sufficient to take $m_0 \, b_0
\simeq 70$ to account for the ratio O$\left( 10^{-\,16} \right)$
between the observed electroweak and gravitational
scales.\footnote{One would expect the inverse of $b_0$ to be naturally
of the order of all the other mass scales, $b_0^{-\,1} \simeq m_0
\simeq M_p \,$. The quantity $m_0 \, b_0 \simeq 70$ needed in the RS
proposal is however a consistent improvement with respect to the ratio
between the electroweak and the Planck scale. Moreover, one may hope
that this value can emerge naturally from the stabilization mechanism
of the dilaton.}
We now want to understand the behavior of the system when matter is
included on the two branes, that is when the energy--momentum tensor
of the two branes is of the form
\begin{eqnarray}
&& \left( T_A^B \right)_{\mathrm brane} = \frac{\delta \left( y \right)}{b_0} \, {\mathrm diag} \left( V_0 + \rho_0 \,,\, V_0 - p_0 \,,\, V_0 - p_0 \,,\, V_0 - p_0 \,,\, 0 \right) + \\
&& + \frac{\delta \left( y - 1/2 \right)}{b_0} \, {\mathrm diag} \left( V_{1/2} + \rho_{1/2} \,,\, V_{1/2} - p_{1/2} \,,\, V_{1/2} - p_{1/2} \,,\, V_{1/2} - p_{1/2} \,,\, 0 \right) \;\;, \nonumber
\end{eqnarray}
where $V_{0\,,\,1/2}$ are the cosmological constants previously defined, while $\rho_i$ and $p_i$ are, respectively, the density and pressure of matter on the two branes with equation of state $p_i = w_i \, \rho_i \;$ ($\,i=0\,,\,1/2\,$). In ref. \cite{CGRT} the system has been solved at first order in $\rho_i$ and $p_i\,$ for the special case $w_0 = w_{1/2} = -1\,$. In the present work we provide {\it exact} solutions of the whole system for {\it arbitrary} equations of state on the two branes. To do this, we make use of the results achieved in ref. \cite{KKOP} in the case of a five dimensional space with one single brane.
First of all, we integrate the Einstein equation (\ref{Einstein05}). This is solved either for $\dot{a}=0$ (in this case one recovers a class of static solutions including the RS solution (\ref{statsol})), or for:
\begin{equation}
n (y,t) = \lambda (t) \dot{a} (y,t) \;\;. \label{relationna}
\end{equation}
This relation introduces an unknown function of time only, and
considerably simplifies the remaining equations. Note that we have a complete freedom in the choice of $\lambda\,$, since different $\lambda$'s correspond to different definitions of the time variable.
By inserting eq. (\ref{relationna}) into eq. (\ref{Einstein00}), 
we can eliminate the time-derivatives, and we obtain
a simple second-order differential equation for $a^2$:
\begin{equation}
\left( a^2 (y,t) \right)^{''} - 4 m_0^2 b_0^2~a^2 (y,t) = 
\frac{2 b_0^2}{\lambda(t)^2}.
\end{equation}
This equation has a solution:
\begin{equation}
a^2 \left( t \,,\, y \right) = a_0^2 \left( t \right) \omega_0^2 \left( y \right) + a_{1/2}^2 \left( t \right) \omega_{1/2}^2 \left( y \right)
+ \frac{\omega_0^2 \left( y \right) + \omega_{1/2}^2 \left( y \right) - 1}{2 m_0^2 \lambda \left( t \right)^2} \;\;,
\label{asol}
\end{equation}
where
\begin{eqnarray}
\omega_0^2 \left( y \right) &=& \cosh \left( 2 \, m_0 \, b_0 \, | y | \right) - \frac{C_0}{S_0} \, \sinh \left( 2 \, m_0 \, b_0 \, | y | \right) \;\;, \nonumber \\
\omega_{1/2}^2 \left( y \right) &=& \frac{{\mathrm sinh} \left( 2 \, m_0 \, b_0 \, | y | \right)}{S_0} \;\;,
\label{omegas}
\end{eqnarray}
with $C_0 \equiv \cosh \left( m_0 \, b_0 \right)$ and $S_0 \equiv \sinh \left( m_0 \, b_0 \right)\,$.
Eq. (\ref{asol}) relates the value of $a \left( t \,,\, y \right)$ in
the whole space to the values on the two branes $a_0 \left( t \right)
\equiv a \left( t \,,\, 0 \right) $ and $a_{1/2} \left( t \right)
\equiv a \left( t \,,\, 1/2 \right) \,$. These two unknown
time--dependent functions are determined below.
Rather than the last remaining non--trivial equation associated to the component (\ref{Einsteinii}) of the Einstein tensor, we consider -- in strict analogy to what is customarily done in conventional FRW cosmology -- the equation associated to the identity \footnote{To achieve this identity, the expression (\ref{relationna}) must be used.}
\begin{equation} 
G_{ii}=
-
\frac{a}{\dot{a}} 
\frac{d}{d\,t} \left\{ 
\frac{a^2}{3n^2}
G_{00}
\right\}
- 
\frac{a^2}{3n^2}
G_{00}.
\end{equation}
Substituting in this expression the components of the Einstein tensor with the corresponding components of the energy--momentum tensor, we get an expression which is trivially satisfied in the bulk, while on the two branes it reduces to
\begin{equation}
\dot{\rho}_i + 3 \frac{\dot{a}_i}{a_i} \, (\rho_i + p_i) = 0 \;,\;\;i=0\,,\,1/2 \;\;.
\label{conservation}
\end{equation}
These two equations are nothing but the energy--conservation law in the two branes and they are identical to the energy--conservation law of standard FRW cosmology.

We finally have to determine the functions $a_0 \left( t \right)$ and $a_{1/2}\left( t \right)$ appearing in expression (\ref{asol}). This can be done by solving eq. (\ref{Einstein00}) across the two branes. As shown in the work \cite{BDL}, 
one can put this last step in form of ``jump conditions'' which relate the discontinuity of $n'$ and $a'$ to the delta--like source $\left( T_A^B \right)_{{\mathrm brane}} \,$. From the symmetry $y \leftrightarrow -y$, we can write the ``jump conditions'' in the form:
\begin{eqnarray}
\frac{a' (0,t)}{a (0,t)} &=& - \frac{\kappa^2}{6} b_0 \left(
V_0+\rho_0 \right),
\nonumber \\
\frac{n' (0,t)}{n (0,t)} &=& \frac{\kappa^2}{6} b_0 \left[ 2 \left( V_0+\rho_0 \right) + 3 \left( -V_0+p_0 \right) \right], 
\nonumber \\
\frac{a' (\frac{1}{2},t)}{a (\frac{1}{2},t)} &=& \frac{\kappa^2}{6}
b_0 \left( V_{1/2}+\rho_{1/2} \right),
\nonumber \\
\frac{n' (\frac{1}{2},t)}{n (\frac{1}{2},t)} &=& - \frac{\kappa^2}{6}
b_0 \left[ 2 \left( V_{1/2}+\rho_{1/2} \right) + 3 \left( -V_{1/2}+p_{1/2} \right) \right] \;\;. \label{jumps}
\end{eqnarray}
These equations lead to the following system for $a_0$, $a_{1/2}$:
\begin{eqnarray}
\left[ 1 + \frac{\kappa^2 \, \rho_0}{6\,m_0} - \frac{C_0}{S_0} \right] \, a_0^2 + \frac{a_{1/2}^2}{S_0} &=& \frac{C_0-1}{2\,m_0^2\,\lambda^2 \, S_0} \nonumber \;\;, \\
\frac{a_0^2}{S_0} + \left[ -\,1 + \frac{\kappa^2 \, \rho_{1/2}}{6\,m_0} - \frac{C_0}{S_0} \right] \, a_{1/2}^2 &=& \frac{C_0 - 1}{2\,m_0^2\,\lambda^2 \, S_0} \;\;.
\label{system}
\end{eqnarray}
As expected, the system admits no solution in absence of matter on the two branes, $\rho_0 = \rho_{1/2} = 0 \;$. Indeed, for this choice one recovers the static RS solution, which is not accounted for by the relation (\ref{relationna}). When matter is instead included, the system (\ref{system}) gives the solutions:
\begin{eqnarray}
a_0^{-\,2} \lambda^{-\,2} &=&
\frac{\kappa^2\,m_0}{3 \left( 1-\Omega_0^2 \right)}
\frac
{
\rho_0 + \Omega_0^4 \rho_{1/2} - \frac{\kappa^2}{12 \, m_0}
\left( 1- \Omega_0^4 \right) \rho_0 \rho_{1/2}
}
{
1- \left( 1-\Omega_0^2 \right) \frac{\kappa^2 \, \rho_{1/2}}{12 \, m_0}
}
\;\;, \label{a0} \\
a_{1/2}^{-2} \lambda^{-2} &=&
\frac{\kappa^2\,m_0}{3 \left( 1-\Omega_0^2 \right)}
~\frac{1}{\Omega_0^2}~
\frac
{
\rho_0 + \Omega_0^4 \rho_{1/2} - \frac{\kappa^2}{12 \, m_0}
\left( 1- \Omega_0^4 \right) \rho_0 \rho_{1/2}
}
{
1 - (\Omega_0^{-\,2}-1) \frac{\kappa^2 \, \rho_0}{12 \, m_0}
} \;\;. \label{a12}
\end{eqnarray}
Since $\lambda = n_0 / \dot{a}_0 = n_{1/2} / \dot{a}_{1/2} \,$, we can interpret these equations as the expansion laws of the two branes. As we will see below, eqs. (\ref{conservation}), (\ref{a0}), and (\ref{a12}) give standard FRW evolution on both branes at low energy.

\section{The effective four dimensional action}

We can gain some insight on the cosmology of the RS model by integrating the whole action over the extra dimension $y\,$. In doing so, we make use of the result (\ref{asol}). Our goal is to get an effective four dimensional action which describes the evolution of the scale factors $a_0 \left( t \right) \;,\; a_{1/2} \left( t \right)$ on the two branes.

We first focus on the ``purely gravitational'' five dimensional action, that is we integrate the RS action in the absence of matter on the two branes. The latter will be considered eventually when we deal with the equations of motion. Our starting point is thus:
\begin{eqnarray}
S&=& - \int\,d^4x\,dy\,\sqrt{g} \left[ \frac{R}{2\,\kappa^2} +
\Lambda + \frac{\delta\left(y\right)}{b_0} V_0 +\frac{\delta\left(y-1/2\right)}{b_0} V_{1/2} \right]
\label{action} \\
&=& - \frac{1}{2\,\kappa^2} \int d^4x\,dy \sqrt{g}\, \left[\,R-12\,m_0^2+12\,m_0\,\left(\frac{\delta\left(y\right)}{b_0}-\frac{\delta\left(y-1/2\right)}{b_0}\right)\right]\;, 
\nonumber
\end{eqnarray}
with the full (five dimensional) curvature scalar given by:
\begin{equation}
R = 6 n^{-2} \left[ \frac{\dot{n}}{n} \frac{\dot{a}}{a}
- \frac{\ddot{a}}{a} - \left( \frac{\dot{a}}{a} \right)^2 \right]
+ 2 b_0^{-2} \left[ \frac{n''}{n} + 3 \frac{n'}{n} \frac{a'}{a}
+ 3 \frac{a''}{a} + 3 \left( \frac{a'}{a} \right)^2 \right] \;.
\end{equation}
Since we are interested in the evolution of the two four dimensional branes, we rewrite $n \left( t \,,\,y \right)$ and $a \left( t \,,\,y \right)$ by making use of the results of the previous section, eqs. (\ref{relationna}) and (\ref{asol}).
It is then convenient to write $\sqrt{g} R$ and $\sqrt{g}$ in terms of $a^2$ and $\lambda$:
\begin{eqnarray}
\sqrt{g} R &=& \frac{6 b_0}{\lambda} \left(
\frac{\dot \lambda}{\lambda} a^2 - \frac{1}{2} \, \frac{d\,a^2}{d t} \right)
+ \frac{\lambda}{2 b_0} \left[ \frac{d  \left( a^4 \right)''}{d t} - 3
\left( a^2 \right)' \frac{d \left( a^2 \right)'}{d t} \right] \;\;, \nonumber \\
\sqrt{g} &=& \frac{\lambda\,b_0}{4} \frac{d \, a^4}{d t} \;\;.
\end{eqnarray}
With all these considerations,\footnote{The calculation can be further simplified by noticing that, from the periodicity imposed in the extra space, the integral of a derivative of any function of $y$ vanishes.}the integral over $y$ of the action (\ref{action}) gives:
\begin{eqnarray}
S&=&-\frac{1}{2\,\kappa^2}\int d^4x \left\{\frac{1-\Omega_0^2}{m_0}\, \frac{1}{1+\Omega_0^2}\left[ \frac{6}{\lambda} \left(\frac{\dot{\lambda}}{\lambda}\, \left(a_0^2+a_{1/2}^2\right)-  \left(a_0\,\dot{a}_0+a_{1/2}\,\dot{a}_{1/2}\right) \right) \right]\,+\right.\nonumber\\
&+&\left.\,\frac{24\,m_0}{1-\Omega_0^4}\,\lambda\,\left[\Omega_0^2\,\left(a_0\,\dot{a}_0\,a_{1/2}^2+a_{1/2}\,\dot{a}_{1/2}\,a_0^2\right)\,-\,\left(\Omega_0^4\,a_0^3\,\dot{a}_0+a_{1/2}^3\,\dot{a}_{1/2}\right)\right]\right\} \,.
\end{eqnarray}
By substituting $\lambda=n_0/\dot{a}_0=n_{1/2}/\dot{a}_{1/2}$ in the last expression\footnote{In this way, we substitute $\lambda \left( t \right)$ with the two degrees of freedom $n_0 \left( t \right)$ and $n_{1/2} \left( t \right) \,$. The equations of motion of the effective four dimensional theory have thus to be supported by the constraint $n_0/\dot{a}_0=n_{1/2}/\dot{a}_{1/2}\,$. This relation cannot be obtained from the action (\ref{effact}), since it is linked to the equation $G_{05}=0$ that has no counterpart in the four dimensional effective theory.} we get:
\begin{eqnarray} \label{effact}
S= &-&\frac{M_p^2}{2 (1+\Omega_0^2)}\,\int d^4x \left\{n_0\,a_0^3\, \frac{6}{n_0^2} \left[ \frac{\dot{n}_0}{n_0} \frac{\dot{a}_0}{a_0}
- \frac{\ddot{a}_0}{a_0} - \left( \frac{\dot{a}_0}{a_0} \right)^2 \right]+\right. \\ &&\hspace{3cm}+\left. n_{1/2}\,a_{1/2}^3\,\frac{6}{n_{1/2}^2} \left[ \frac{\dot{n}_{1/2}}{n_{1/2}} \frac{\dot{a}_{1/2}}{a_{1/2}}
- \frac{\ddot{a}_{1/2}}{a_{1/2}} - \left( \frac{\dot{a}_{1/2}}{a_{1/2}} \right)^2 \right]\right. + \nonumber \\
&+& \left. \frac{24\,m_0^2}{\left(1-\Omega_0^2\right)^2} 
\left[\Omega_0^2\,\left(a_0\,n_0\,a_{1/2}^2+a_{1/2}\,n_{1/2}\,a_0^2\right)-\left(\Omega_0^4\,a_0^3\,n_0+a_{1/2}^3\,n_{1/2}\right)\right] \right\}\,\,.\nonumber
\end{eqnarray}

As we will discuss in more detail in the next section, in the low
energy limit the equality $a_{1/2}(t) = \Omega_0\, a_{0}(t)$ and the
related one $n_{1/2}(t)=\Omega_0\, n_{0}(t)$ hold. As a consequence,
the expansion rates of the two branes are identical and the above
action rewrites in the standard FRW form:
\begin{equation}  
S= -\frac{M_p^2}{2}\,\int d^4x\, \bar{n}\, \bar{a}^3\, \frac{6}{\bar{n}^2} \left[ \frac{\dot{\bar{n}}}{\bar{n}}\, \frac{\dot{\bar{a}}}{\bar{a}}
- \frac{\ddot{\bar{a}}}{\bar{a}} - \left( \frac{\dot{\bar{a}}}{\bar{a}} \right)^2 \right]\,\,,
\label{estat}
\end{equation}
where $\bar{a} \equiv a_0 = \Omega_0^{-\,1} \, a_{1/2} \;\;,\;\; \bar{n} \equiv n_0 = \Omega_0^{-\,1} \, n_{1/2} \;$.

{}From the effective action (\ref{effact}) we notice that the entire five dimensional system can be (``holographically'') expressed in terms of the physics that takes place on the boundaries at $y=0$ and $y=1/2$ of the extra space. Notice also that the last term in the action (\ref{effact}) couples the metrics of the two walls.

Going back to the four-dimensional action (\ref{effact}), and
including also matter on the two walls, we obtain the equations of
motion:
\begin{eqnarray}
&&\frac{\dot{a}_0^2}{n_0^2\,a_0^2}=\frac{1+\Omega_0^2}{3\,M_p^2}\,\rho_0+\frac{4\,m_0^2}{\left(1-\Omega_0^2\right)^2}\, \Omega_0^2\, \left( \frac{a_{1/2}^2}{a_0^2}-\Omega_0^2\right)\nonumber\\
&&\frac{\dot{a}_{1/2}^2}{n_{1/2}^2\,a_{1/2}^2}=\frac{1+\Omega_0^2}{3\,M_p^2}\,\rho_{1/2}+\frac{4\,m_0^2}{\left(1-\Omega_0^2\right)^2}\, \Omega_0^2\, \left( \frac{a_0^2}{a_{1/2}^2}-\frac{1}{\Omega_0^2}\right)\,\,\,,
\label{effeq}
\end{eqnarray}
in addition to the relations which give energy conservation on the two branes, eqs. (\ref{conservation}).
We notice that, in the limit $\rho_0\rightarrow 0$, $\rho_{1/2}\rightarrow 0\,$, the only solution of the above equations is the static RS solution $a_{1/2}=\Omega_0 a_{0}$. Moreover, one can verify that eqs. (\ref{effeq}) are equivalent to the equations (\ref{a0}) and (\ref{a12}) obtained from the five dimensional theory.

\section{FRW evolution at low energy}

Before interpreting the four-dimensional effective theory found above, we come back to the static RS case. We recall that in \cite{RS} the four-dimensional metric $\bar{g}_{\mu \nu}$ on both branes is defined as:
\begin{equation}
\bar{g}_{\mu \nu} = n(y)^{-2} g_{\mu \nu} \;.
\end{equation}
The goal of this redefinition is to achieve Minkowski metric on both
branes, in order to gain a simple physical interpretation of the system.
An analogous procedure has to be applied also in the general case with
matter on the two branes.

Generally speaking, multiplying the metric by an overall function $f$ is
not equivalent to a change of the coordinate system. Thus, to have
canonical normalization of the fields, the function $f$ has to be absorbed
by a redefinition of the fields themselves. In order to preserve the
equations of motion of the fields, we see that we cannot choose $f$ to depend on the coordinates $t$ and $x\,$, but it can be at most a function of $y\,$.

In analogy with what was done in the static case,
we now wonder whether it is possible to rewrite the first four components of
the five-dimensional metric in the form: 
\begin{equation}
g_{\mu \nu} \left( t \,,\,y \right) = f \left( y \right) \,\bar{g}_{\mu \nu} \left( t \right) \;,
\label{fact}
\end{equation}
with $\bar{g}_{\mu \nu}$ of the standard FRW form
diag$(1,-\bar{a}^2,-\bar{a}^2,-\bar{a}^2)$.  This requires the ratio
$n/a$ to be independent on $y\,$, that is $a'/a=n'/n$ for every value
of $y\,$. From the ``jump conditions'' (\ref{jumps}) we see that this
implies $\rho+p=0$ and, consequently, $\dot{\rho}=0$ on the two
branes. In other words, the above factorization is possible only if
the two branes contain exclusively cosmological constants (in
particular this is the case for the static RS solution).

Anyhow, it is natural to expect that condition (\ref{fact}) is
approximately recovered when the matter on the two branes has a
sufficiently low energy density. This can be understood from the
results of the previous sections. From eqs. (\ref{a0}) and (\ref{a12})
we have:
\begin{equation}
\frac{a_{1/2}^2}{a_0^2} = \Omega_0^2 \; \Frac{1 - \frac{\kappa^2 \left( 1 - \Omega_0^2 \right)}{12\,m_0\,\Omega_0^2} \, \rho_0}{1 - \frac{\kappa^2 \left( 1 - \Omega_0^2 \right)}{12\,m_0} \, \rho_{1/2}} \;.
\label{cond}
\end{equation}
If $\rho_0$ and $\rho_{1/2}$ are sufficiently small, the scale factors
of the two branes are (approximately) proportional \footnote{Notice
that this relation holds exactly in the static RS case.} by the
constant factor $\Omega_0\,$. Since $n \left( y \,,\, t \right) =
\lambda \left( t \right) \dot{a} \left(y \,,\, t \right)$, we have
also $n_{1/2} \left( t \right) = \Omega_0 n_{0} \left( t \right)
\,$. In particular, the ratio $n/a$ is (approximately) independent on
$y\,$. \footnote{From eqs. (\ref{asol}), (\ref{a0}), and (\ref{a12}), it is indeed possible to show that, in the low energy limit, the quantity $n'/n - a'/a$ is of the same order as $a_{1/2} / \left( \Omega_0 a_0 \right) - 1 \,$.} 

In this low-energy limit, we can thus define the four-dimensional effective
theory just as in the static RS model. First, we can choose the time
coordinate so that $n_0$ and $n_{1/2}$ are simultaneously
time-independent. This is equivalent to setting $\lambda(t)=\lambda_0
/ \dot{a}_{1/2}(t)$, where $\lambda_0$ is an arbitrary factor. Then,
we recover eq.(\ref{fact}) with:
\begin{equation}
f(0) = \lambda_0^2~\Omega_0^{-2}, \qquad 
f(1/2) = \lambda_0^2, \qquad \bar{a}=\lambda_0^{-1} \Omega_0 a_0
=\lambda_0^{-1} a_{1/2}~.
\label{lambda0}
\end{equation}
We can now use the freedom to fix the time coordinate, and choose a
particular value of $\lambda_0$. Choosing $\lambda_0=\Omega_0$ we
recover, in the limit $\rho_0 \,,\, \rho_{1/2} \rightarrow 0\,$, the
static RS solutions as presented in \cite{RS}. With this choice, the scale factor
of the effective metrics reads
$\bar{a}=a_0=\Omega_0^{-1}~a_{1/2}$.

We can identify the five-dimensional quantities with those measured at low energy in our brane: 
\begin{itemize}
\item
the fields must be redefined by a factor $\Omega_0$. So, for instance,
the observed density is $\bar{\rho}_{1/2} = \Omega_0^4~\rho_{1/2}$. On
the other brane the canonically normalized density reads:
$\bar{\rho}_{0} = \rho_{0} \,$.
\item
the total four dimensional effective action (\ref{effact}) acquires the form of the standard FRW action in terms of the scale factor $\bar{a}\,$, see eq. (\ref{estat}). 
\item
The Hubble parameter of the low energy theory is given by $\dot{\bar{a}}/\bar{a}\,$. From both eq. (\ref{a0}) and eq. (\ref{a12}) we get the standard evolution law:
\begin{equation}
H^2 = \left( \frac{\dot{\bar{a}}}{\bar{a}} \right)^2 \simeq \frac{1}{3 \, M_p^2} \: \left( \bar{\rho}_0 + \bar{\rho}_{1/2} \right) \;\;,
\end{equation}
while from eqs. (\ref{conservation}) we recover
\begin{eqnarray}
\dot{\bar{\rho}} &+& 3 \frac{\dot{\bar{a}}}{\bar{a}} \, (\bar{\rho} + \bar{p}) = 0 \;\;, \nonumber \\
\bar{\rho} &\equiv& \bar{\rho}_0 + \bar{\rho}_{1/2} \;\;,\;\;
\bar{p} \equiv \bar{p}_0 + \bar{p}_{1/2} \;\;.
\end{eqnarray}
\end{itemize}

Some considerations are in order. First, we would like to emphasize
that at low energy, from the point of view of observers on both
branes, the effective theory leads exactly to a standard
four-dimensional FRW Universe. This follows from the fact that the
standard Friedmann law is recovered, and that the energy densities on
both branes scale with the same Hubble parameter.  In particular, for
what concerns observers on our brane, the matter on the $0$--brane is
regarded as dark matter \cite{CGRT} that would completely escape any
direct or indirect experimental detection (apart of course from its
gravitational interactions). The gravitational effect of the matter on
the $0$--brane is not suppressed by powers of $\Omega_0\,$, as it is
the case for $\bar{\rho}_{1/2}\,$. Since the only natural mass scale
of the model is the Planck scale, $\bar{\rho}_0$ must hence be
fine--tuned to small values not to conflict with observations (see
the next section).

Second, we remark that some care has to be paid in the interpretation
of the physically observable quantities in the low energy effective
theory. For instance, the alternative choice $\lambda_0 =1$
\cite{KIM} in eqs. (\ref{relationna}) and (\ref{lambda0}) is not
compatible with the identification of $\bar{\rho}_{1/2} = \Omega_0^4
\rho_{1/2}$ as the observed energy density on our brane. This would
lead to a misinterpretation of the expansion laws of the two branes.

Then, in order to put quantitative limits on the validity of the low energy theory, we rewrite eq. (\ref{cond}) in terms of the observed matter densities: \footnote{We use $m_0 \simeq \kappa^{-\,2/3} \simeq M_p$ and $M_p \, \Omega_0 \simeq$ TeV.}
\begin{equation}
\frac{a_{1/2}^2}{a_0^2} = \Omega_0^2 \; \Frac{1 - \frac{\bar{\rho}_0}{10\,M_p^2\,{\mathrm TeV}^2}}{1 - \frac{\bar{\rho}_{1/2}}{10\,{\mathrm TeV}^4}} \;\;.
\label{cond2}
\end{equation}
We see that the low energy approximation is valid as long as the observed matter densities satisfy the bounds:
\begin{equation}
\bar{\rho}_0 \ll 10\,M_p^2\:{\mathrm TeV}^2 \;\;,\;\; \bar{\rho}_{1/2} \ll 10\:{\mathrm TeV}^4 \,. \label{lowenrcond}
\end{equation}
Finally, we would like to comment on Planck mass in the RS model. At
low energy, there are two possible ways to define it, one related to
the five--dimensional expansion, and one from the four dimensional
effective action. These two
definitions are called, respectively, {\it local} and {\it global} in
ref. \cite{MPP}. We see that indeed the values of $M_p$ obtained
with these two definitions coincide once all the quantities in the four
dimensional action are properly identified.

\section{Corrections to Standard Cosmology at high energy}

We now focus on the equations of motion when the low-energy conditions
(\ref{lowenrcond}) are not fulfilled anymore. From what we said in the
previous section, it is clear that in this regime it is not possible
any longer to have a simple interpretation of the effective four
dimensional action in terms of observable quantities. However, this
is not important, because we make measurements only today, in the
low-energy limit. So, it is legitimate to study the evolution of the
system at high energy (eqs. (\ref{conservation}), (\ref{a0}), and
(\ref{a12})), and then make contact with the quantities that we
observe today.\footnote{This remark should be important, for instance,
when looking at cosmological perturbations in the early Universe. In
this section, we derive only the evolution equations of the homogeneous
background. When studying the perturbations, one should keep in mind
that a full five-dimensional description is required at high energy.}

We keep the previous definitions of $\bar{\rho}_i$, $\bar{p}_i$,
$M_P$, and the choice $\lambda=\Omega_0/\dot{a}_{1/2}$, so that
eqs.(\ref{a12}) and (\ref{conservation}) rewrite:
\begin{eqnarray}
\left( \frac{\dot{a}_{1/2}}{a_{1/2}} \right)^2 &=&
\frac{1}{3 \, M_p^2} \,
\frac
{
\bar{\rho}_0 + \bar{\rho}_{1/2} - \frac{\kappa^2}{12 \, m_0}
\left( \Omega_0^{-\,2} - \Omega_0^2 \right) \bar{\rho}_0 \bar{\rho}_{1/2}
}
{
1 - (\Omega_0^{-\,2}-1) \frac{\kappa^2 \, \bar{\rho}_0}{12 \, m_0}
} \;\;, \\
\dot{\bar{\rho}}_{1/2} &+& 3 \,
\frac{\dot{a}_{1/2}}{a_{1/2}}\,
\left(\bar{\rho}_{1/2} + \bar{p}_{1/2}\right) = 0 \;\;.
\end{eqnarray}
With our ansatz for $\lambda(t)$, the warp factor on our brane is
constant. So, all the Euler-Lagrange equations on our brane are the
same at high and low energy (i.e., they remain exactly identical to
the standard equations of physics in four dimensions)\footnote{Since
the freedom in choosing $\lambda(t)$ is equivalent to the freedom in
fixing the time coordinate, it is obvious from general relativity
principles that all physical results would not be affected by
another choice of $\lambda(t)$, with the correct low-energy behavior
$\lambda(t) \rightarrow \lambda_0 / \dot{a}_{1/2}(t)$. It is
meaningless to wonder which choice of $\lambda(t)$ makes sense
physically at high energy, since contact with observations is only
made at low energy. So, it is sufficient to give the set of equations
that follows from the simplest choice for $\lambda(t)$.}. In order to
close the differential system, we need an equation of evolution for
$\bar{\rho}_0$. It is obtained from
eqs. (\ref{conservation}), (\ref{a0}), and (\ref{a12}):
\begin{equation}
\dot{\bar{\rho}}_0 = - 3 \frac{\dot{a}_{1/2}}{a_{1/2}} (\bar{\rho}_0 + \bar{p}_0)
\left( 1 - \frac{3(\bar{\rho}_0 + \bar{p}_0)}
{2(\frac{12 m_0}{\kappa^2} \frac{\Omega_0^2}{1-\Omega_0^2}-\bar{\rho}_0)}
\right)
\left( 1 - \frac{3(\bar{\rho}_{1/2} + \bar{p}_{1/2})}
{2(\frac{12 m_0}{\kappa^2} \frac{\Omega_0^4}{1-\Omega_0^2}-\bar{\rho}_{1/2})}
\right)^{-1}
\end{equation}
The differences between the evolution equations for $\bar{\rho}_0$ and
$\bar{\rho}_{1/2}$ (i.e., the terms in the parentheses) show
explicitly that, at high energy, $\bar{\rho}_0$ is not equivalent to
dark matter in our brane.

Since it is assumed that $m_0 \simeq \kappa^{-\,2/3} \simeq M_p$ and
that $\Omega_0 \, M_p \simeq$ TeV~, the above equations can be cast in
the more transparent form :
\begin{eqnarray} \label{fineq}
&&\left(\frac{\dot{a}_{1/2}}{a_{1/2}} \right)^2 
= \frac{\bar{\rho}_{1/2}}{3 \, M_p^2} \,
\left(
1 + \frac{\bar{\rho}_0}{\bar{\rho}_{1/2}} - \frac{\bar{\rho}_0}{10 \, 
M_p^2 \, {\mathrm TeV}^2}
\right)
\left(
1 - \frac{\bar{\rho}_0}{10 \, M_p^2 \, {\mathrm TeV}^2}
\right)^{-1} \;\;, \label{deviation} \\
&&\dot{\bar{\rho}}_0 = -3 \frac{\dot{a}_{1/2}}{a_{1/2}} 
(\bar{\rho}_0 + \bar{p}_0)
\left( 1 - \frac
{3(\bar{\rho}_0 + \bar{p}_0)}
{2(10 M_P^2 \, {\mathrm TeV}^2 - \bar{\rho}_0 )}
\right)
\left( 1 - \frac
{3(\bar{\rho}_{1/2} + \bar{p}_{1/2})}
{2(10\, {\mathrm TeV}^4 - \bar{\rho}_{1/2})}
\right)^{-1}\;\;. \nonumber
\end{eqnarray}
We now discuss the implications of these equations for the cosmological
evolution in the early Universe.

First of all, it is worth noticing that the above equations (\ref{fineq})
encounter a singularity when $\bar{\rho}_0 \simeq 10 M_P^2 \, {\mathrm
TeV}^2\,$. It may be possible that the presence of such singularity puts a
limit on the theory, at least as long as the dilaton is assumed to be stabilized.
However, as we will show later, phenomenological bounds from primordial
nucleosynthesis indicate that this limit is hardly reached for $\bar{\rho}_{1/2}
\leq {\mathrm TeV}^4\,$.

In the regime of validity of the low-energy effective theory,
$\bar{\rho}_{0}$ behaves as ordinary dark matter in our brane. So, the
constraints that we usually have for dark matter apply to it.
Although in principle we cannot say much about the physics on the
0-brane (in particular ``non--standard'' equations of state may be
expected), we assume for simplicity that $\bar{\rho}_0$ can be
decomposed into a constant term $\bar{\rho}_0^{\Lambda}$ 
($w_0=-1$), plus matter $\bar{\rho}_0^m$ and
radiation $\bar{\rho}_0^{r}$ components (with $w_0=0,1/3$).

For what concerns the constant component, the sum of the
cosmological terms $\bar{\rho}_0^{\Lambda}$ and
$\bar{\rho}_{1/2}^{\Lambda}$ is bounded by the current value of the
critical density, which is of order $10^{-123}~M_P^4$. So,
the amount of fine-tuning required here is the same
as in usual 4-dimensional theories:
\begin{equation}
\bar{\rho}_0^{\Lambda}+\bar{\rho}_{1/2}^{\Lambda} = \rho_0^{\Lambda} + \Omega_0^4\, \rho_{1/2}^{\Lambda} \leq 
10^{-123}~M_P^4.
\end{equation}
The matter and radiation components also have to be fine-tuned to
small values.  The best current constraint on the radiation density
$\bar{\rho}_0^r$ comes from nucleosynthesis: since the observed
abundances of light elements are only compatible with an effective
number of neutrinos $N_{eff}=3 \pm 1$, we see that $\bar{\rho}_0^r$ is
bounded by the density of one family of relativistic neutrinos. The
matter density $\bar{\rho}_0^m$ is obviously bounded by the value of
the critical density today.  So, in the five-dimensional theory, both
$\rho_0^r$ and $\rho_0^m$ have to be fine-tuned to $\sim \Omega_0^4
~\rho_{1/2}^r$ and $\sim \Omega_0^4~\rho_{1/2}^m$, while one may
naively expect $\rho_0 \sim \rho_{1/2}$ in the early Universe. 

Without the knowledge of the behavior of the RS model at high energy, one may
have hoped that corrections to the standard Friedmann law could have solved
this problem. For example, starting from $\rho_0 \sim \rho_{1/2}$ at high
energy, the equations of motion of the system could have naturally lead to
$\rho_0 \ll \rho_{1/2}$ at temperatures of the order of the one at which primordial nucleosynthesis occurred. Our
analysis shows that this is not the case. Indeed, let us assume
$\bar{\rho}_{1/2} \sim \bar{\rho}_0$ at the nucleosynthesis scale ($\bar{\rho}_i \sim
{\mathrm MeV}^4$) and let us consider the behavior of the system when it was
close to the natural cut--off $\bar{\rho}_{1/2} \sim {\mathrm TeV}^4\,$.
Significant deviations from the standard evolution are expected if at that
epoch the  energy $\bar{\rho}_0$ was almost of order $M_p^2 \, {\mathrm TeV}^2$
[see eq. (\ref{fineq})]. Going backwards in time, $\bar{\rho}_0$ can increase
relatively to $\bar{\rho}_{1/2}$ if $w_0>w_{1/2}\,$. However, assuming
radiation domination on our brane above the nucleosynthesis scale, the above requirement
can be met only for $w_0 \geq 2 \,$, which does not seem to be a realistic
possibility.

A possible solution of the problem of the fine--tuning of
$\bar{\rho}_0$ may arise from the stabilization mechanism for the
dilaton, especially if it occurs at (relatively) low energy. Other possibilities are briefly discussed in ref. \cite{CGRT}.

\section{Conclusions}

In this work we have studied the cosmological evolution of the Randall--Sundrum model \cite{RS} with matter on the two branes. We have first provided exact analytical solutions for the model, valid for arbitrary equations of state of the matter on the branes. By integrating the system over the extra dimension $y\,$, we have then obtained an effective four dimensional action.

These results can be used to investigate the physical behavior of the model. We have seen that at low energy the branes expand with the same rate and 
standard FRW cosmology is recovered on both of them. From our point of view, matter on the other brane is seen as dark matter.

When one goes to higher energies, the physical interpretation of the system in terms of four dimensional quantities becomes less clear, since the $y$ dependence cannot be factorized away from the five dimensional metric, as occurs in the static case and at low energy. However, the results presented in the first part of this work hold also in the high energy regime. So one can still look at the evolution of the system at high energy, and then make contact with the quantities that we presently observe in the low energy theory.

Denoting with $\bar{\rho}_{1/2}$ (respect. $\bar{\rho}_0$) the observed energy density on our (respect. on the other) brane, we have found that the low effective (FRW) theory is valid for
\begin{equation}
\bar{\rho}_{1/2} \ll {\mathrm TeV}^4 \;\;,\;\; \bar{\rho}_0 \ll {\mathrm TeV}^2 \, M_p^2 \;\;.
\end{equation}

As a consequence, corrections to standard cosmology cannot be found at energies much smaller than these values. This is the main result of our paper, since it extends the previous analysis \cite{CGRT} valid only at first order in constant energy densities.

The main motivation for the RS scenario is that it has only one fundamental
scale $M_p\,$. From the definition of the physically observable quantities,
this would suggest $\bar{\rho}_{1/2} \sim {\mathrm TeV}^4$ and $\bar{\rho}_0
\sim M_p^4$ to be the most natural initial values for the matter
densities on the two branes. While the first value appears to be acceptable,
our analysis shows that the equation of motion of the system are meaningful
only up to $\bar{\rho}_0 \sim 10 \, {\mathrm TeV}^2 \, M_p^2 \,$. Moreover, the
phenomenological bound $\bar{\rho}_0 < \bar{\rho}_{1/2}$ which has to be
imposed from primordial nucleosynthesis on, forces $\rho_0$ to be negligible
with respect to $\rho_{1/2}$ in the five dimensional theory. We have seen that
the evolution of the system does not lead to this hierarchy at the
nucleosynthesis period unless $\bar{\rho}_0$ has a non--standard equation of state.

\section{Acknowledgments}
We would like to thank Jussi Kalkkinen for useful discussions. J.~L.~ and S.~P.~ are supported by INFN and by the European Commission under TMR network grant ERBFMRXCT960090.


\begin{thebibliography}{99}

\bibitem{ADD}
N.~Arkani-Hamed, S.~Dimopoulos and G.~Dvali,
Phys.\ Lett.\  {\bf B429} (1998) 263
[hep-ph/9803315].

\bibitem{RS}
L.~Randall and R.~Sundrum,
Phys.\ Rev.\ Lett.\  {\bf 83} (1999) 3370
[hep-ph/9905221].

\bibitem{BDL}
P.~Bin\'etruy, C.~Deffayet and D.~Langlois,
Nucl.\ Phys.\  {\bf B565} (2000) 269
[hep-th/9905012].

\bibitem{CGS}
J.~M.~Cline, C.~Grojean and G.~Servant,
Phys.\ Rev.\ Lett.\  {\bf 83} (1999) 4245
[hep-ph/9906523].

\bibitem{BDEL}
P.~Bin\'etruy, C.~Deffayet, U.~Ellwanger and D.~Langlois,
Phys.\ Lett.\ {\bf B477} (2000) 285
[hep-th/9910219].


\bibitem{SMS}
T.~Shiromizu, K.~Maeda and M.~Sasaki,
gr-qc/9910076.


\bibitem{FTW}
E.~E.~Flanagan, S.~H.~Tye and I.~Wasserman,
hep-ph/9910498.

\bibitem{CGRT}
C.~Cs\'aki, M.~Graesser, L.~Randall and J.~Terning,
hep-ph/9911406.

\bibitem{KKOP}
P.~Kanti, I.~I.~Kogan, K.~A.~Olive and M.~Pospelov,
hep-ph/9912266.

\bibitem{KIM}
H.~B.~Kim,
hep-th/0001209.

\bibitem{GW}
W.~D.~Goldberger and M.~B.~Wise,
Phys.\ Rev.\ Lett.\  {\bf 83} (1999) 4922
[hep-ph/9907447].

\bibitem{MPP}
R.~N.~Mohapatra, A.~P\'erez-Lorenzana and C.~A.~de Sousa Pires,
hep-ph/0003328.

\end{thebibliography}
\end{document}